\documentclass[reprint,onecolumn,amsmath,amssymb,jap,aip]{revtex4-1}
\usepackage{graphicx}
\usepackage{dcolumn}
\usepackage{bm}
\usepackage{epsfig}
\usepackage{mathptmx}
\usepackage{times}
\usepackage{amsmath}
\usepackage{amssymb}
\usepackage{dcolumn}
\usepackage{units}
\usepackage{upgreek}
\usepackage{tabularx}
\usepackage{todonotes}
\usepackage{threeparttable}
\usepackage[space]{grffile}
\usepackage[normalem]{ulem}

\usepackage{subfig}
\usepackage{multirow}
\usepackage{array}
\usepackage{xcolor}
\usepackage{soul}
\usepackage[]{caption}
\usepackage{siunitx}
\begin{document}

\title{Projected mushroom type phase-change memory}
\author{Syed Ghazi Sarwat} \email{ghs@zurich.ibm.com} \affiliation{IBM Research -- Europe, S\"{a}umerstrasse 4, 8803 R\"{u}schlikon, Switzerland}
\author{Timothy M.Philip} \affiliation{IBM Research AI Hardware Center-- Albany,  12203 NY, USA}
\author{Ching-Tzu Chen} \affiliation{IBM Research AI Hardware Center-- Albany,  12203 NY, USA}
\author{Benedikt Kersting} \affiliation{IBM Research -- Europe, S\"{a}umerstrasse 4, 8803 R\"{u}schlikon, Switzerland}
\author{Robert L Bruce} \affiliation{IBM Research--  Yorktown Heights,  10598 NY, USA}
\author{Cheng-Wei Cheng} \affiliation{IBM Research--  Yorktown Heights,  10598 NY, USA}
\author{Ning Li} \affiliation{IBM Research--  Yorktown Heights,  10598 NY, USA}
\author{Nicole Saulnier} \affiliation{IBM Research AI Hardware Center-- Albany,  12203 NY, USA}
\author{Matthew BrightSky} \affiliation{IBM Research--  Yorktown Heights,  10598 NY, USA}
\author {Abu Sebastian} \email{ase@zurich.ibm.com} \affiliation{IBM Research -- Europe, S\"{a}umerstrasse 4, 8803 R\"{u}schlikon, Switzerland} 

\maketitle


\begin{flushleft}
\end{flushleft}

\noindent{\textbf{Phase-change memory devices have found applications in in-memory computing where the physical attributes of these devices are exploited to compute in place without the need to shuttle data between memory and processing units. However, non-idealities such as temporal variations in the electrical resistance have a detrimental impact on the achievable computational precision. To address this, a promising approach is projecting the phase configuration of phase change material onto some stable element within the device. Here we investigate the projection mechanism in a prominent phase-change memory device architecture, namely mushroom-type phase-change memory. Using nanoscale projected Ge$_{2}$Sb$_{2}$Te$_{5}$ devices we study the key attributes of state-dependent resistance, drift coefficients, and phase configurations, and using them reveal how these devices fundamentally work.}} \\

\noindent{Phase change memory (PCM) devices can not only store data in their adjustable physical state but also use that data to perform computations by modifying an externally applied signal. This attribute has enabled their use in the emerging computational scheme of in-memory computing\cite{Y2018zidanNatureElectronicsReview,Y2018ielminiNatureElectronics,Y2019mutluMandM,Y2020AbuNatNano,Y2018sebastianJAP,Y2011kuzumNanoletters,Y2018ielminiNatureElectronics,Y2018yuProcIEEE}. However, PCM in-memory computing continue to be based on more conventional architectures of PCMs, which were developed for binary data storage\cite{Y2010Wong,Y2011Xiong,Y2016Arasu,Y2017Ghazi} and not computation. Important device non-idealities from the intrinsic material physics of phase-change materials\cite{Y2020ManuelJPD,Y2009nardonePRB,Y2016legalloESSDERC,Y2010ielminiAPL,Y2019carboniAEM,Sebastian2019,Y2017GhaziMST} thus arise and reduce the numerical precision achievable with this technology. Both materials and device engineering have been proposed as solutions\cite{Y2018LIunderstandingdrfit,Y2019PCMHeterostructure, y2013xiong,Y2019thermalshen,Y2021heterogeneouslyyang} to minimizing these non-idealities. Device engineering invokes novel cell architectures, including the concept of projected PCM\cite{Y2015koelmansNatComm,Y2020BenediktScientificReports,Y2013kimIEDM,Y2016kimIEDM,Y2010ProjectedLinerPatent}}. The concept decouples the readout characteristics of the device from the noisy electrical properties of the phase-change material. This is realized using an electrically conducting material, called the projection layer (liner) placed in parallel to the phase-change material. Thus, in effect, the device read-out characteristics become dictated by the properties of the projection layer. \\

The understanding of ‘projection’ has, however, remained limited to lateral-type architectures, which are not yet established for scaling-up. Of interest, therefore, are vertical devices with established materials and processing methods, that can be densely integrated and provide an optimized balance of key device properties, such as the mushroom-type PCM devices\cite{Y2021BOBBYIRPS,Y2010Wong,Y2010Burr,Y2010PhaseChangeMemory,Y2010ProjectedLinerPatent}.  In this article, we investigate the device characteristics of projected mushroom-type PCM (PMPCM) devices, and study the unique characteristics that distinguish them from their traditional unprojected counterparts. To that end, we construct an analytical  model, and experimentally verify it using nanoscale devices projected to different extents. We specifically focus on the relevant properties of state-dependent resistance drift and temperature sensitivity, using which, we identify and discuss some guidelines for the device design of PMPCM. 

\section*{Model of a projected mushroom-type PCM device}
\begin{figure}[h!]
    \includegraphics[width=\textwidth]{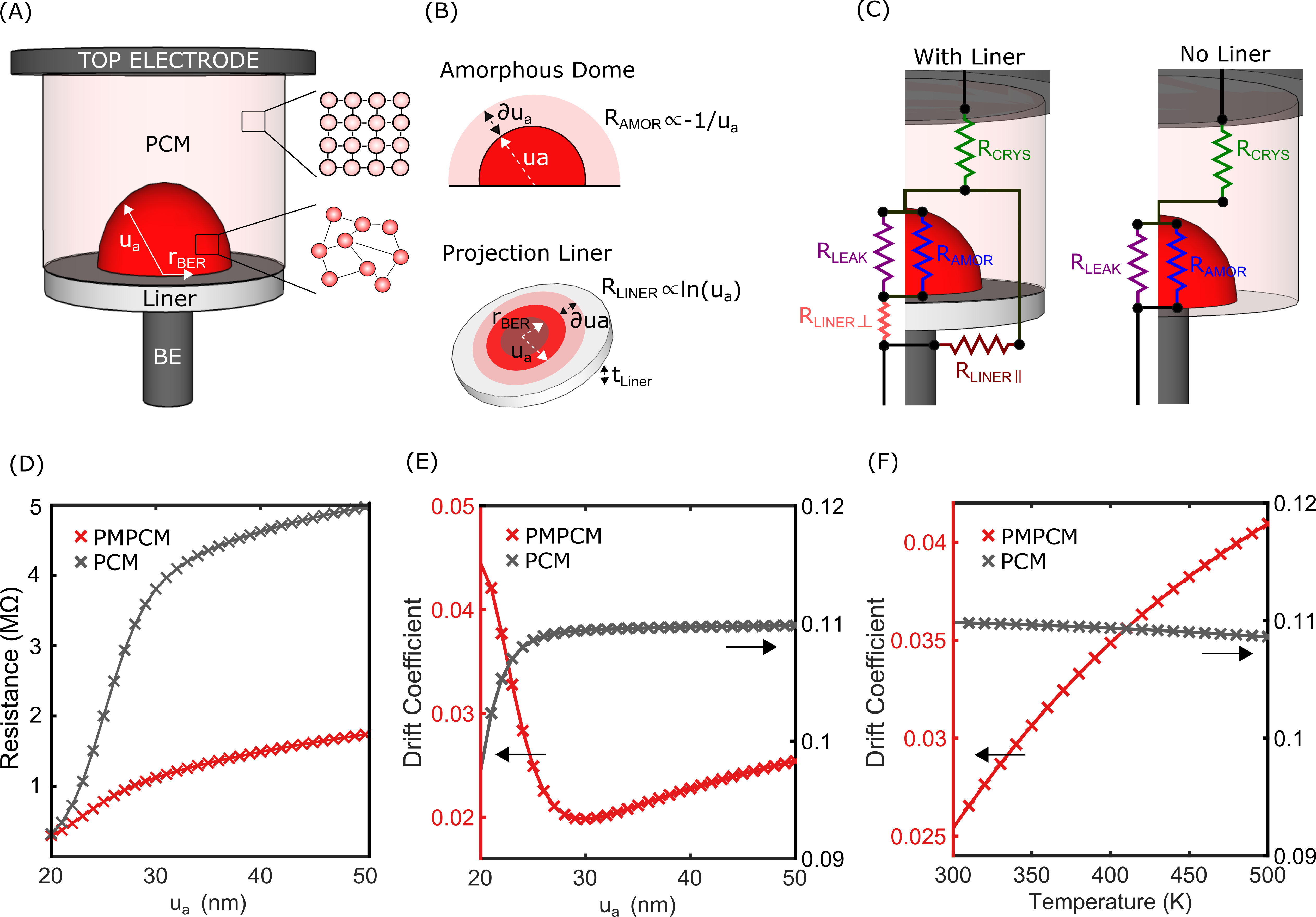}
    \caption{\textbf{Projected mushroom-type PCM.} (A) Schematic illustration of a PMPCM device, with the key components highlighted. (B) Illustrations showing the geometrical approximations for the amorphous phase-change material segment and the liner. (C) An electrical circuit model highlighting the different lumped resistors that comprise a projected and unprojected mushroom-type device. A projected device has an additional component, that is parallel to the amorphous phase-change material dome.  (D) Simulated resistance vs amorphous dome size ($u_{\text{a}}$) in a projected and an unprojected device. (E) Simulated drift coefficients as a function of $u_{\text{a}}$ for projected and unprojected devices. (F) Simulated drifts coefficients scaling of a RESET state in projected and unprojected devices with ambient temperature. Figures D to F highlight the key distinguishing features between projected and unprojected mushroom-type devices.}
    \label{fig:1}
\end{figure}

\noindent Figure \ref{fig:1}A illustrates a sketch of a PMPCM device. The device consists of a layer of phase-change material sandwiched between a top (TE) and a narrower bottom (BE) electrode. Uniquely placed between the bottom electrode and the phase-change material is a thin film, which acts as the liner that spans the width and breadth of the device in dimensions. The resistance state of the device is dictated by the phase configuration, or the amount and geometry of amorphous volume (colored in red) in an otherwise crystalline phase-change material (colored in pink). A short, high current pulse (RESET), when applied to a device, amorphizes a fraction of the crystalline phase-change material in the region abutting the bottom electrode of radius $r_{\text{BE}}$. The amorphous volume brings the PCM device to a high-resistance state, where the resistance magnitude is dictated by the radius of this volume, which we refer to as $u_{\text{a}}$. If one breaks down a PMPCM device in the RESET state into its lumped resistive components, it will comprise: $R_{\text{Amor}}$ (resistance to current from amorphous volume), $R_{\text{Crys}}$ (resistance from the remaining crystalline volume), $R_{\text{Leak}}$ (resistance from some shunt resistor), and $R_{\text{Liner}}$ (resistance from the liner). $R_{\text{Liner}}$ can further be broken into two resistors, namely $R_{\text{Liner}\perp}$ and $R_{\text{Liner}\parallel}$. $R_{\text{Liner}\perp}$ is the resistance presented by the liner to the current that flows perpendicularly through it (this can be seen as the current that flows perpendicular to $r_{\text{BE}}$), and $R_{\text{Liner}\parallel}$ is the resistance presented by the liner to the current that flows into it (this can be seen as the current that flows parallels to $r_{\text{BE}}$).  If an approximation is then made for the amorphous volume to have shape of a hemispherical dome, and the liner as a cylindrical disc\cite{Y2009Bipin,Y1967Holm,Y2010AbuAPL} (see Figure \ref{fig:1}B), the $u_{\text{a}}$ dependence of $R_{\text{Amor}}$, $R_{\text{Crys}}$ and $R_{\text{Liner}}$ on geometrical arguments can be expressed by the terms in the square brackets of equations  

\begin{eqnarray}\label{eq6}
R_{\text{Amor}}(u_{\text{a}},T,t) &=& \left[ \rho_{\text{Amor,o}}\times \left(\frac{1}{8 \times r_{\text{BE}}}+\frac{1}{2 \times \pi} \times \left(\frac{1}{r_{\text{BE}}}-\frac{1}{u_{\text{a}}}\right)\right)\right] \times \exp\left(\frac{E_{\text{Amor}}}{k_{\text{b}}T}\right) \times \left(\frac{t}{t_{0}}\right)^{\upsilon_\text{Amor}}\\\label{eq1}
R_{\text{Crys}}(u_{\text{a}},T,t) &=& \left[\frac{\rho_{\text{Crys,o}}}{2 \times \pi} \times \left(\frac{1}{u_{\text{a}}}-\frac{1}{t_{\text{PCM}}}\right)\right]\times \exp\left(\frac{E_{\text{Crys}}}{k_{\text{b}}T}\right) \times \left(\frac{t}{t_{0}}\right)^{\upsilon_{\text{Crys}}}\\\label{eq2}
R_{\text{Liner}\parallel}(u_{\text{a}},T,t) &=& \left[\frac{\rho_{\text{Liner,o}\parallel}}{2 \pi \times t_{\text{PCM}}} \times (\ln(u_{\text{a}})-\ln(r_{\text{BE}}))\right] \times \exp\left(\frac{E_{\text{Liner}}}{k_{\text{b}}T}\right) \times \left(\frac{t}{t_{0}}\right)^{\upsilon_{\text{Liner}}}\\\label{eq3}
R_{\text{Liner}\perp}(u_{\text{a}},T,t) &=& \left[\frac{\alpha \times t_{\text{Liner}}}{\pi \times r_{\text{BE}}^{2}}\right] \times \exp\left(\frac{E_{\text{Liner}}}{k_{\text{b}}T}\right)  \times \left(\frac{t}{t_{0}}\right)^{\upsilon_{\text{Liner}}}\\\label{eq4}
R_{\text{Leak}}(u_{\text{a}},T,t) &=& \left[\frac{\rho_{\text{Leak,o}} \times u_{\text{a}}}{\pi \times r_{\text{c}}^{2}}\right] \times \exp\left(\frac{E_{\text{Leak}}}{k_{\text{b}}T}\right) \times \left(\frac{t}{t_{0}}\right)^{\upsilon_{\text{Leak}}}
\end{eqnarray}

where, $\rho_{\text{Amor,o}}, \rho_{\text{Crys,o}}, \rho_{\text{Liner,o}}$, and $\rho_{\text{Leak,o}}$ are the electrical resistivities of the sub-scripted components; they are temperature independent Arrhenius prefactors at time instance $t_{\text{0}}$. $t_{\text{\text{PCM}}}$ and $t_{\text{Liner}}$ are the thickness of the phase-change material, and the liner respectively, and other symbols have their previously defined meanings. Note that these equations are only valid for the regime where $u_{a}\geq r_{BE}$. $t_{\text{Liner}}$ in our investigation are sub-\unit[10]{nm}, and for such a film thinness, we invoke directional anisotropy\cite{Y2007Anisotropy2,Y2007Anisotropy} in $\rho_{\text{Liner,o}}$. It is expressed as $\rho_{\text{Liner,o}\perp}$ and $\rho_{\text{Liner,o}\parallel}$, with reference to whether the current flows parallel or perpendicularly to $t_{\text{Liner}}$, respectively. Equation 1 also takes into account the spreading resistance (first part in the square bracket) that models the current flow deviating from a straight course\cite{Y1967Holm}. Both these effects are also expected for equation 4, and we take them into account using $\alpha=\frac{\rho_{\text{Liner,o}\perp}}{n}$, where \textit{n} is a positive rational number.  The shunt resistance $R_{\text{Leak}}$ takes into account any electrical conducting pathways that may exist in the amorphous phase-change material dome and is a proxy for the shunt (parasitic) resistor used commonly to account for conductive defects in the modeling of solar cells\cite{Y2020PVEducation,Y2012Shunt,Y2009BENGHANEM}. We model $R_{\text{Leak}}$ in equation 5 as a cylindrical volume of radius $r_{\text{c}}$, and height $u_{\text{a}}$, where $r_{\text{c}}\propto \exp(-u_{\text{a}})$. $R_{\text{Leak}}$ is parallel to $R_{\text{Amor}}$, therefore the effective resistance of the amorphous dome can be expressed as $ R_{\text{Amor} \parallel \text{Leak}} = \frac{R_{\text{Amor}} \times R_{\text{Leak}}} {R_{\text{Amor}}+R_{\text{Leak}}} $. \\

The thermally activated charge transport (\textit{T} dependence) and resistance drift (\textit{t} dependence) are accounted in the above equations by expressions after the square brackets, respectively. $E_{\text{Amor}}, E_{\text{Crys}}, E_{\text{Liner}}$, and $E_{\text{Leak}}$ are the activation energies and $\nu_{\text{Leak}},  \nu_{\text{Crys}},\nu_{\text{Crys}}$ and $\nu_{\text{Leak}}$ are the drift coefficients of the sub-scripted components. $t$ is the time elapsed after $t_{0}$, $k_{\text{b}}$ and $T$ are the Boltzmann constant (in eV) and ambient temperature (in K), respectively. Using these lumped resistors we frame an equivalent electrical circuit of a mushroom-type device, with and without a liner. This is sketched in Figure \ref{fig:1}C. The key difference between the two device types is that in a PMPCM device an additional resistor (which is the liner) appears in parallel to the amorphous phase-change material volume. The total device resistance of a standard unprojected and a PMPCM device can be therefore expressed by the following equations

\begin{eqnarray}\label{eq5}
R_{\text{Total(unprojected)}} &=& R_{\text{Amor} \parallel \text{Leak}}+R_{\text{Crys}}\\\label{eq6} 
R_{\text{Total(PMPCM)}} &=&  \frac{(R_{\text{Liner}\perp}+R_{\text{Amor} \parallel\text{Leak}}) \times R_{\text{Liner}\parallel}} {(R_{\text{Liner}\perp}+R_{\text{Amor} \parallel \text{Leak}})+R_{\text{Liner}\parallel}}+R_{\text{Crys}}\label{eq7}
\end{eqnarray}

$R_{\text{Liner}\perp}$ in eqn.\ref{eq7} can be also considered as a component that is in series with both $R_{\text{Amor} \parallel \text{Leak}}$ and $R_{\text{Crys}}$ (see section S1). The essential benefit of a PMPCM device is that the liner decouples information storage from the information read-out. This occurs due to the parallel circuit configuration, since the majority of current flows `preferentially' through $R_{\text{Liner}\parallel}$, instead of traversing through $R_{\text{Amor}}$. To gain an initial overview of the key differences in the device metrics with and without the liner, we plug the required material and device parameters into our circuit model. We plot the resistance and drift coefficient as a function $u_{\text{a}}$ and $T$, as illustrated in Figure \ref{fig:1}D-F, where $u_{\text{a}}$ spans from \unit[50]{nm} $\geq u_{\text{a}} \geq r_{\text{BE}}$ (\unit[20]{nm}), and the $T$ from \unit[300]{K}-\unit[500]{K}. Figure 1D suggests that the resistance of an unprojected device scales more dramatically than an equivalent PMPCM device, and for the same $u_{\text{a}}$ an unprojected device is more resistive than the PMPCM device. In device terms, this would imply that an unprojected device provides a larger memory window than an effectively projecting PMPCM device.  Figure 1E illustrates that in unprojected device, the drift coefficients are large, and more or less invariant to $u_{\text{a}}$. The PMPCM device however shows much reduced drift coefficients, and the $u_{\text{a}}$ dependency can be segmented into three distinct regions. At small $u_{\text{a}}$ (region 1) the drift coefficients are large and scale  inversely with $u_{\text{a}}$, for intermediate $u_{\text{a}}$ (region 2) the drift coefficients plateau at their smallest values, and at large $u_{\text{a}}$ (region 3), the drift coefficients increase with $u_{\text{a}}$. Figure 1F shows the drift coefficients in both device types as a function of temperature for an amorphous dome of size $u_{\text{a}}$ (\unit[50]{nm}). The drift coefficients in an unprojected device do not change with the ambient temperature, but in the PMPCM device become temperature dependent.  In what follows, we will explore these trends experimentally.

\section*{Experimental Validation}
\subsection*{Activation energies, phase configuration and amorphous dome size }

\begin{figure}[h!]
    \includegraphics[width=0.85\textwidth]{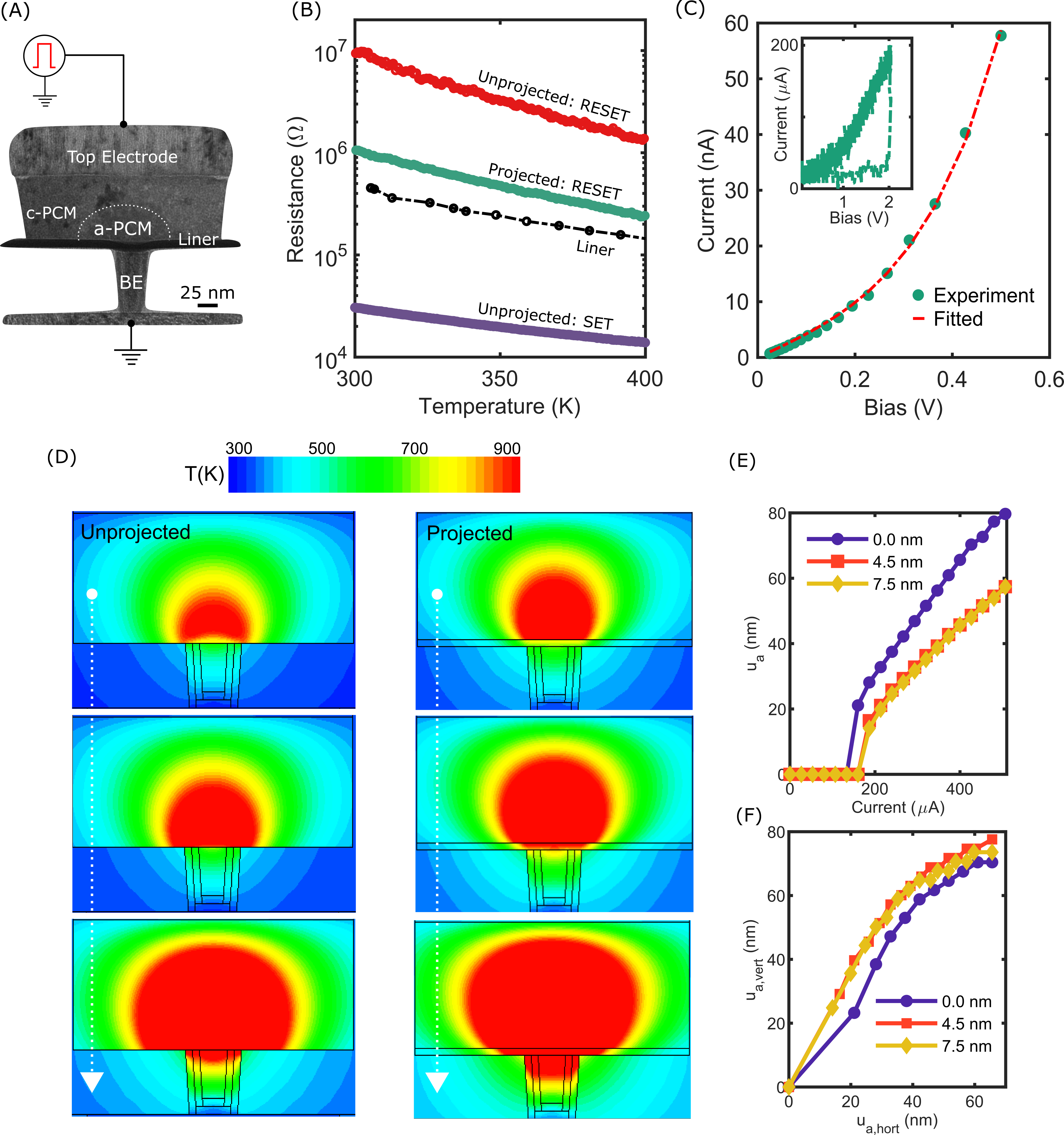}
    \caption{\textbf{Device parameters.} (A) A transmission electron micrograph of a doped GST based PMPCM device, with the different parts labeled. (B) Resistance vs temperature measurements of the SET and RESET states of an unprojected device, a liner film and RESET state of a PMPCM device. From the Arrhenius equation fitting of these measurements, the activation energies for thermally activated transport in the different components are determined. (C) A current-voltage trace of a RESET state in an unprojected device. The data is fitted with a modified Poole-Frenkel equation, as highlighted the in red. Inset illustrates the threshold switching induced crystallization of this state. (D) Electro-thermal simulations of a standard mushroom-type device under increasing programming current. The melting temperature of dGST is 900 K, thus the region shaded in red is an approximation of $u_{\text{a}}$. The right plot is a similar simulation on a PMPCM device with a liner thickness of 4.5 nm. (E) Simulated programming curves of the unprojected and PMPCM devices with different liner thicknesses. (F) A plot comparing the asymmetry in the geometry of the amorphous dome between unprojected and PMPCM devices. }
    \label{fig:2}
\end{figure}

\noindent A key parameter needed to verify the circuit model is the charge-transport activation energies ($E$) of the different device components making a PMPCM device. To obtain these we measure unprojected and PMPCM devices. Figure \ref{fig:2}A is a transmission electron micrograph of a typical PMPCM device based on doped-Ge$_{2}$Sb$_{2}$Te$_{5}$ (d-GST) phase-change material. The different components are indicated. We fabricated PMPCM devices with liner thicknesses of 7.5$\pm$1.0 nm and 4.5$\pm$1.0 nm (see Methods and section S2-S3). To extract activation energies, we perform RT (resistance as a function of temperature) measurements on the crystalline (SET) and amorphous (RESET) states of an unprojected device, an \unit[8]{nm} thick blanket film of the liner, and a RESET state of a PMPCM device (see Figure \ref{fig:2}B-C). We extract $E_{\text{Amor}}$, $E_{\text{Crys}}$ and $E_{\text{Liner}}$, by fitting the RT data to the Arrhenius equation (see Methods). From these measurements we deduce $E_{\text{Amor}}$ = \unit[0.21]{eV}, $E_{\text{Crys}}$ = \unit[0.08]{eV} and $E_{\text{Liner}}$ = \unit[0.12]{eV}. A PMPCM device has  $E_{\text{Amor effective}}$ = \unit[0.15]{eV}, suggesting that the liner decreases the device's temperature sensitivity as would be expected due to projection.  \\

Another important metric that requires verification is the geometry of the amorphous volume in PMPCM devices. In the circuit model we approximated this to be hemisphere. To study this, we simulated the steady-state behavior of mushroom devices using a custom electro-thermal finite-element method (FEM) model (see Methods). Figure \ref{fig:2}D illustrates the calculated temperature profile of both an unprojected mushroom-type device and a projected device with a 4.5 nm thick projection liner. Note that the amorphous volume evolves as a hemispherical dome\cite{athmanathan2016multi,Y2019Neumann,Y2018Thermoelectric} in both device types. And, although the dome size is not dramatically altered by the presence of the liner, we do observe that the center of the hot spot in PMPCM device rises higher relative to that of the unprojected device. This occurs because the liner more efficiently mediates the heat transfer from the phase-change material to the bottom electrode, thus resulting to a puckered dome. Figure \ref{fig:2}E shows the amorphous thickness, $u_{\text{a}}$, for increasing programming currents in unprojected, and projected devices (with liner thicknesses of 4.5 nm, and 7.5 nm). We observe that at any given programming current, the effective dome size is smaller in PMPCM devices, which can be attributed to the puckering effect. We also note a negligible dependence of dome size on the liner thickness. Figure \ref{fig:2}F plots the vertical dimension of the amorphous dome, $u_{\text{a,vert}}$  against the horizontal dimension of the dome, $u_{\text{a,hort}}$. The higher hotspot in projected devices can more easily be observed here as the vertical extent in the projected devices is larger than the horizontal extent. 
Our FEM simulations therefore show that although subtle changes in the amorphous phase configuration do occur in projected devices, the projection liner does not significantly alter the amorphous phase generation in these devices- as such, the hemispherical amorphous dome approximation in PMPCM devices is justified, even when volumetric thermoelectric effects, and thermal and electrical boundary resistances are included (see section S4).\\

The amorphous dome sizes ($u_{\text{a}}$) corresponding to the various programmable resistance states must also be obtained experimentally, in order to make fair comparisons. To extract $u_{\text{a}}$, we make use of the standard field-voltage relation $u_{\text{a}} = \frac{V_{\text{th}}}{F_{\text{th}}}$. Here, $F_{\text{th}}$ is the threshold-switching field unique to d-GST and $V_{\text{th}}$ is the threshold-voltage at which the snap-back in the current-voltage measurement occurs. An example measurement is shown in Figure \ref{fig:2}C. Here we perform a current-voltage measurement on an arbitrarily programmed RESET state of an unprojected device, and fit the current-voltage trace to the modified Poole-Frenkel transport model\cite{Y2010papandreouSSE,Iilmininthresholdswitching,Y2008ielminiPRB} (see Methods) to estimate $u_{\text{a}}$. The $V_{\text{th}}$ corresponding to this state is then obtained by applying a long crystallization (SET) pulse (see Figure 2C inset). Through these measurements, we estimate $F_{\text{th}}$ = \unit[54.70]{$MVm^{-1}$} for d-GST (which is similar to GST \cite{Y2010papandreouSSE}). Since $F_{\text{th}}$ is known, $u_{\text{a}}$ of other resistance states can be estimated simply by measuring $V_{\text{th}}$ unique to those states. We use this scheme in the following sections.

\subsection*{State dependent drift coefficients}

\begin{figure}[h!]
    \includegraphics[width=\textwidth]{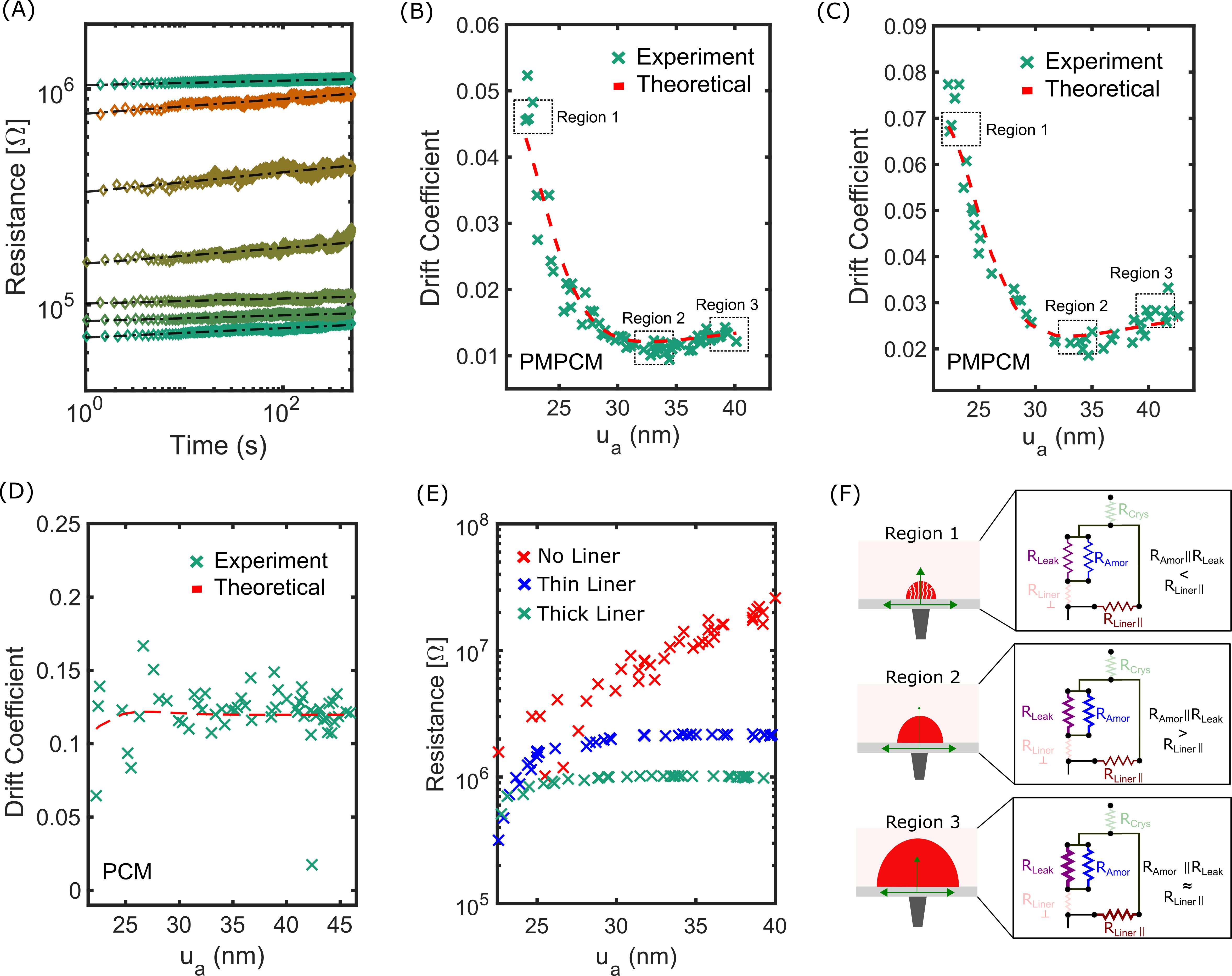}
    \caption{\textbf{State-dependent drift coefficient in projected and unprojected mushroom-type devices.} (A) Resistance vs time plots of many resistance states in a projected device. The data is fitted with the resistance drift equation (black dotted traces). (B) The drift coefficients as a function of $u_{\text{a}}$ in an 7.5 nm thin liner based projected device. The red trace is the fit to the data using the analytical model. (C) Similar measurements as in (B) on a 4.5 nm thin liner based projected device. (D) Drift coefficients vs $u_{\text{a}}$ in an unprojected device. (E) Device resistance as a function of $u_{\text{a}}$ in an unprojected device, projected device with 7.5 nm liner (thick) and with 4.5 nm liner (thin). (F)  Sketches illustrating the mechanisms leading to the state-dependent drift behavior in projected devices.
    \label{fig:4}
    }
\end{figure}

\noindent Having thus obtained the various important device metrics, we now investigate how the drift coefficients ($\nu$) in the unprojected and PMPCM devices scale as a function of $u_{\text{a}}$.  Figure \ref{fig:4}A illustrates resistance vs time measurements of multiple resistance states in a mushroom-type device. By fitting such data with the standard resistance drift equation, we obtain the state-dependent drift coefficients. In Figure \ref{fig:4}B, we plot the drift coefficients of the various unique resistance states in a 7.5 nm liner-based PMPCM device. An equivalent plot for an unprojected device is shown in section S3, from which we extract $\nu_{\text{Amor }}$ and $\nu_{\text{Crys}}$ of d-GST to be $\sim\!0.12$ and $\sim\!0.028$, respectively. Thus, in a PMPCM device, we make two observations (see section S5). First that resistance drift is minimized: for RESET states, drift coefficients are reduced by a factor of 10. This is indicative of good projection efficacy. And second that the drift coefficients scale inversely with $u_{\text{a}}$ for small $u_{\text{a}}$, plateau for an intermediate range of $u_{\text{a}}$, and show a gradual rise for higher $u_{\text{a}}$. In an ideal PMPCM device, the upper limit of drift coefficients is $\leq \nu_{\text{Crys}}$, since majority read current bypasses the amorphous dome ($R_{\text{Amor}}$), and flow through the serial combination of $R_{\text{Liner}}$ and $R_{\text{Crys}}$, where only $R_{\text{Crys}}$ has the tendency to drift. Therefore, the measured high drift coefficients for small $u_{\text{a}}$ indicates that a comparable current also traverses through the amorphous dome, along regions that structurally relax. This observation necessitates presence of a shunt resistor $R_{\text{Leak}}$, which, from a circuit standpoint must be in parallel to $R_{\text{Amor}}$. It is difficult to draw a physical picture of $R_{\text{Leak}}$, although taking ideas from photovoltaics, $R_{\text{Leak}}$ may reflect the many randomly placed conductive pathways bridging the bottom electrode and $R_{\text{Crys}}$ in the device. In an alternate scheme, $R_{\text{Leak}}$ may be representative of extremely fine recrystallized amorphous like grains, at the amorphous-crystalline phase-change material interface (see section S6). \\

We fit the circuit model to this data, and in the spirit of capturing an accurate physical picture, we fit most parameters. The model can adequately trace the data (red trace in Figure \ref{fig:4}B), and the various parameters output by the model's fit are bracketed within the expected margins, which we estimated independently (see Table 1 ($2^{nd}$ and $3^{rd} $columns)). To gain further confidence on the $\nu(u_{\text{a}})$ scaling, we repeated the exact measurements, however on PMPCM devices with a 4.5 nm liner. One such measurement is shown in Figure \ref{fig:4}C. We observe a similar trend in the drift coefficients with respect to $u_{\text{a}}$ but with some differences; namely the drift coefficients peak at higher values, the smallest drift coefficients are $\sim$2x higher than in thicker liner based PMPCM, and the region 3 is more pronounced.  These characteristics indicate reduced projection efficacy in thinner liners based PMPCMs, and are simply a result of the higher sheet resistance of the liner, due to its small physical thinness. When $\nu(u_{\text{a}})$ is fitted to the circuit model, we find an adequate match, and the various parameters lie within the expected margins (see Table 1, $4^{th}$ column). To compare, in Figure \ref{fig:4}D, we plot $\nu(u_{\text{a}})$ in an unprojected PCM device. Notably all resistance-states drift with $\nu_{\text{Amor}}$, and when this experimental data is fitted to the model we observe adequate match (shown as a red trace), and the parameters have expected values (see Table 1, $5^{th}$ column). Such a device behavior can be simply explained by the fact that in an unprojected device, for $u_{\text{a}} \geq r_{\text{BE}}$, the read-current always (and only) traverses through the amorphous dome. Therefore, the drift coefficients can only take a value from a distribution around mean $\nu_{\text{Amor}}$, independent of the $u_{\text{a}}$. The measurement also hints at the nature of $R_{\text{Leak}}$. It suggests that $\nu_{\text{Leak}}$ is bound to $\leq \nu_{\text{Amor}}$; if it were greater, the unprojected devices should have shown higher drift coefficients. In Figure \ref{fig:4}E, we plot the resistance vs $u_{\text{a}}$ of an unprojected device, and PMPCM devices with thinner (4.5 nm), and thicker (7.5 nm) liners. Much in the manner predicted by the circuit model, the unprojected device is more resistive for all $u_{\text{a}}$. This follows equations 1 and 3, where the resistance scales as $\approx \frac{1}{u_{\text{a}}}$ in an unprojected device, whereas in a PMPCM device dominantly by $\approx ln(u_{\text{a}})$. Also observe that within PMPCM devices, the thinner liner devices are more resistive in the RESET states than thicker liner-based, due to the higher $R_{\text{Sheet}}$ of the former. \\

We now discuss the mechanisms that lead to the $\nu(u_{\text{a}})$ dependency in PMPCM devices by sketching the different operational regimes (see Figure \ref{fig:4}F, in which the thinnesses of the individual resistor wriggles encode resistance magnitude). In region 1, where the device is in the regime of a small $u_{\text{a}}$, the effective amorphous dome resistance ($ R_{\text{Amor} \parallel \text{Leak}}$) is comparable to $R_{\text{Liner}\parallel}$ due to the small $R_{\text{Leak}}$. Because these resistive elements are the only two pathways the read-out current (highlighted by green arrows) can take while traversing from the BE to TE, it equally splits. Since a large fraction of the current flows through the amorphous dome, which is the component drifting most in the circuit, the drift coefficients are high. In region 2, or for an intermediate $u_{\text{a}}$, the effective amorphous dome resistance is large compared to $R_{\text{Liner} \parallel}$ due to large $R_{\text{Amor}}$ and $R_{\text{Leak}}$. The implication is that a majority of read-current flows through the liner and not the amorphous dome, thereby yielding small drift coefficients. In region 3, where $u_{\text{a}}$ is large, the difference between the effective amorphous dome resistance and $R_{\text{Liner}\parallel}$ gets smaller through geometrical scaling of $R_{\text{Liner}\parallel}$ and $R_{\text{Amor}}$, such that more and more read-current begins to flow through the amorphous dome. Through a different mechanism, this again leads to a reduction of the projection efficacy. Because this behavior is dictated by $R_{\text{Liner}\parallel}$, thinner liners that have large $R_{\text{Sheet}}$ are expected to show an early onset (at smaller $u_{\text{a}}$) of region 3, as has been experimentally noted.

\subsection*{Temperature dependent drift coefficients}

\begin{figure}[h!]
    \includegraphics[width=0.6\textwidth]{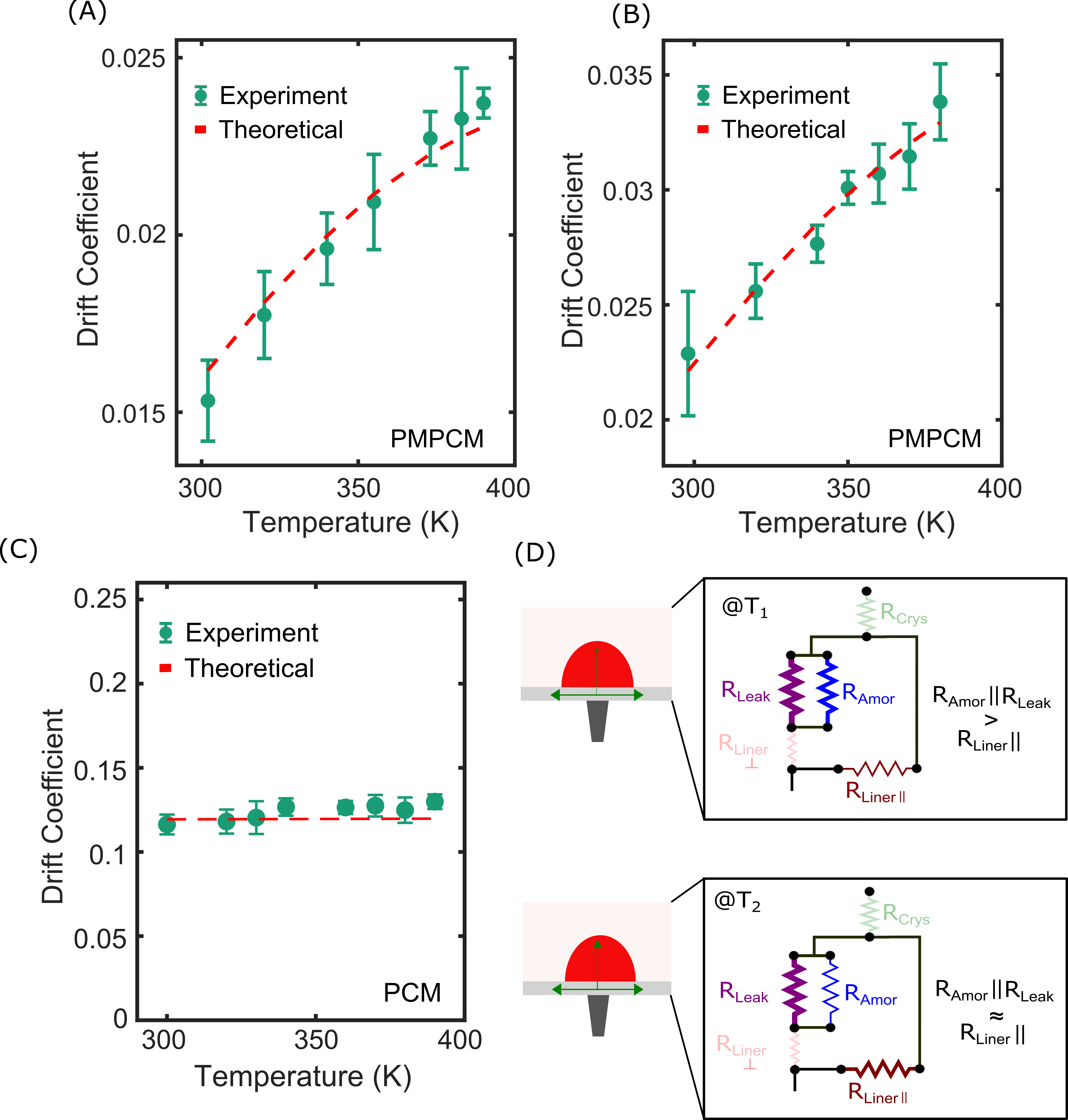}
    \caption{\textbf{Temperature dependent drift coefficient in projected and unprojected mushroom-type devices.} Drift coefficients in an \unit[7.5]{nm} liner based projected mushroom device against ambient temperature, indicating their temperature sensitivity. The red trace shows a fit with the device model. (B) Similar measurements as in (A) on a 4.5 nm liner based projected device. The drift coefficients show similar behavior and temperature sensitivity. (C) The drift coefficients against temperature in an unprojected device. (D) Sketches illustrating the mechanism governing the drift coefficients dependency to temperature in projected devices. }
    \label{fig:5}
\end{figure}

\noindent Ambient temperature is yet another important state-variable that can impact device behavior. We now discuss the temperature dependence of the drift coefficients in unprojected and PMPCM devices. We perform these measurements by creating a RESET state and performing multiple resistance vs time measurements on it at different ambient temperatures (achieved through a heated sample stage). Figure \ref{fig:5}A illustrates one such measurement in a 7.5 nm liner PMPCM device. We observe that the drift coefficients change proportionally with temperature. If a linear approximation is made, the drift coefficient sensitivity to temperature is a $ \sim 0.6\%/^{\circ}C$. We fit the data to the circuit model and note the model to adequately trace the data (see red trace), and all extracted parameters to be in expected range (see Table 1, $6^{th}$ column). We repeat the measurements on 4.5 nm liner-based PMPCM devices (see Figure \ref{fig:5}B). Notably, the drift coefficients show similar temperature dependency, and when the data is fitted to the circuit model, it adequately matches (see red trace). The $\nu(T)$ dependency indicates that the standard resistance drift equation is re-defined for PMPCM devices. This has important implications for in-memory computations since these devices are expected to operate under varying ambient conditions (in automotive, cloud for example). Note, however, that unlike for the dependency on $u_{\text{a}}$, the drift coefficients in both the thinner and thicker liner PMPCM devices scale similarly. This indicates that the dependency arises from changes in the materials' properties (and not geometry). To support these findings, we performed measurements on unprojected devices (see Figure \ref{fig:5}C). Note that the device drifts with a constant drift coefficient $\nu_{\text{Amor}}$ at all temperatures (there is no $T$ dependency). When this data is fitted to the model we find an adequate match (see red trace) and the extracted parameters are bounded within margins (see Table 1, $8^{th}$ column).  \\

The mechanisms that lead to the $\nu(T)$ dependency in the PMPCM devices are discussed in Figure \ref{fig:5}D. We present two cases, one where the device is at room temperature ($T_{1}$) and the other where it is at some elevated temperature ($T_{2}$). At $T_{1}$, the effective dome resistance is larger than $R_{\text{Liner}}$, and therefore a majority of the read-current flows through the non-drifting $R_{\text{Liner}}$ and $R_{\text{Crys}}$. As the ambient temperature increases, $R_{\text{Amor}}$, $R_{\text{Crys}}$, $R_{\text{Leak}}$ and $R_{\text{Liner}}$ decrease -not from changes in the $u_{\text{a}}$, but from the thermally activated charge transport- since both the phase-change material and liner are semiconductors. What determines the $\nu(T)$ dependency however is the extent to which the individual components change relative to each other. A measurement of the temperature sensitivity is the activation energy. In a PMPCM device since $E_{\text{Amor}}$ is greatest, $R_{\text{Amor}}$ drops most dramatically with temperature. Thus, at $T_{2}$, although the resistance of both the amorphous dome and liner drop, the change in resistance of the amorphous dome is larger. This results in the read-current between the parallel pathways to get altered disproportionately, such that more current now traverses through the amorphous dome. Note that at large $u_{\text{a}}$, $R_{\text{Crys}}<<R_{\text{Liner}}$, thus $R_{\text{Crys}}$ can be ignored. \\
Since the drift coefficient of the amorphous dome itself is temperature independent (from Figure \ref{fig:5}C), and the effective dome resistance progressively approaches $R_{\text{Liner}\parallel}$ due to thermally activated charge transport, the device drifts with higher drift coefficients. We additionally verified this by simulating the $\nu({T})$ dependencies for different $E_{\text{Liner}}$ using the circuit model (see section S7). It is observed that as the mismatch between $E_{\text{Liner}}$ and $E_{\text{Amor}}$ grows, the device becomes more temperature-sensitive. When $E_{\text{Liner}}$ equals $E_{\text{Amor}}$, the drift coefficients show least temperature dependency. However, while such a materials science approach reduces the drift’s dependency, it amplifies the device's resistance sensitivity to temperature.  When $E_{\text{Liner}}$ approaches $E_{\text{Amor}}$, the resistance state becomes more and more temperature-sensitive, which can prove detrimental to analog training/inference workloads or multi-level data storage. Thus, while materials can be optimized, adaptive software corrections could also be used (see section S7).

\subsection*{Conclusion}
\noindent{In this paper, we discussed the important material and electrical attributes of projected mushroom-type phase-change memories. Using a combined experimental and theoretical framework, we described how a conductive thin film (liner) when added to a mushroom device modifies the data read-out characteristics. In spite of similar programming and phase configurations to unprojected devices, we found the state-dependent resistance drift and temperature sensitivity to evolve differently in the projected devices. We observed the projection efficacy to be decisively determined by the liner’s physical thickness, and the mismatch of electrical resistivity and the electronic activation energy between the phase-change material and the liner. Our findings provide the first insights into how vertically configured projected memristive devices may fundamentally operate. The results lay a directed framework for building higher performing projected memory for computational-memory applications.} 

\section*{Acknowledgments}

\noindent This work is supported by the IBM Research AI Hardware Center.
We also acknowledge support from the European Research Council through the European Union’s Horizon $2020$ Research and Innovation Program under grant number $682675$. We would like to thank Geoffrey W. Burr and Vijay Narayanan for technical discussions and management support. We thank Linda Rudin for proofreading the manuscript. \\

\section*{Methods}

\noindent{\textbf{Device fabrication and characterization:}} We fabricated Projected PCM devices based on a typical mushroom-type geometry by placing a projection liner between the heater and phase-change material. The devices comprise an 80 nm thick film of a doped-$Ge_{2}Sb_{2}Te_{5}$ (d-GST) phase-change material, a sub-10 nm thick film of metal-nitride ($M_{x}N_{y}$) liner, bottom and top $M_{x}N_{y}$ electrodes, where the bottom electrode radius is 19 nm. The device is 540 nm wide, and equally broad. All material depositions were carried out using sputter depositions. The devices were made on an 8-inch wafer, and we observed spatial nonuniformity's in the thicknesses of the different films. This was noted using ellipsometry measurements, and we took these variations into account in our analysis. We characterized the retention, threshold voltage, SET speed and resistance drift properties of the projected and unprojected devices. The results are discussed in the supporting information. We calibrated the FEM model using the activation energy measurements on unprojected devices. For simulations, the current flow in the device is modeled by enforcing the continuity equation at each mesh point, using the expression $\nabla.J=0$ and $J=-\sigma \nabla V$, where $J$ is current density, $\sigma$ is electrical conductivity, and $V$ is the voltage. The crystalline phase-change material conductivity follows a Poole-Frenkel-type activated-transport model. To capture the self-heating that governs the operation of these devices, the continuity equation is solved simultaneously with the energy conservation equation $-\nabla.(k\nabla T)=J.(\sigma^{-1}J)$, where $k$ is the thermal conductivity of the material and $T$ is the local temperature in the device. The thermal properties of the materials here are assumed to follow both the Weidemann-Franz electronic relationship. All parameters including electrical and thermal interface (boundary) resistances are taken from measurements and tabulated in section S4. In our analytical modelling a device without the liner comprises only $R_{\text{Amor}}$, $R_{\text{Crys}}$ and $R_{\text{Leak}}$. When $u_{\text{a}}=r_{\text{BE}}$ in both device types, the BE can be referred to as fully covered or blocked, in which case a fraction of current traversing from BE to TE must always flow through the resistive amorphous volume. In this case, the resistances of the BE and TE to current are insignificant and can be ignored. In our fitting to the temperature dependent drift coefficient, similar to resistivity anisotropy, we also considered anisotropy in $E_{\text{Liner}}$. From fitting we observed $E_{\text{Liner}\perp}<0.5E_{\text{Liner}\parallel}$, regardless data could also be fitted without anisotropy. \\

\noindent{\textbf{Device measurements:}} The electrical measurements were performed in a custom-built probe station. The temperature of the chuck was measured using a Lake Shore Si DT-670B-CU-HT diode. Devices were electrically contacted with high-frequency Cascade Microtech Dual-Z GSSG probes. DC measurements of the device state were performed with a Keithley 2600 System SourceMeter. AC signals were applied to the device with an Agilent 81150 A pulse function arbitrary generator. A Tektronix oscilloscope (DPO5104) recorded the voltage pulses applied to and transmitted by the device. Switching between the circuit for DC and AC measurements was achieved with mechanical relays. Note that in our measurements all READ operations were performed at 0.2 V. We extract $E_{\text{Amor}}$, $E_{\text{Crys}}$ and $E_{\text{Liner}}$ by fitting the resistance vs temperature data to the Arrhenius equation of the form
$E=k_{\text{b}} \times T\times \ln\left(\frac{R_{\text{T}}}{R_{\infty}}\right)$ where, $R_{T}$ is resistance at temperature $T$, $R_{\infty}$ is the device resistance at $\infty$ $K$ and other symbols have their usual meanings. We obtain the state-dependent $u_{\text{a}}$ by using the field equation. As a starting point, from a known $V_{\text{th}}$ and $u_{\text{a}}$, we extract $F_{\text{th}}$. For this we measure the current-voltage trace of a RESET state and fit the trace to the modified Poole-Frenkel transport model $I=2qAN_{tot} \times \frac{\Delta z}{\tau_{0}}\times \exp\left(-\frac{E_{C}-E_{F}}{k_{\text{b}}T}\right) \times \sinh\left(\frac{qV}{k_{\text{b}}T}\frac{\Delta z}{2u_{\text{a}}}\right)$ where, $I$ is the current (in A), $V$ is the voltage (in V), $q$ is the electrical charge, $A$ is the contact area (in $m^{2}$), $N_{tot}$ is the trap density (in $m^{-3}$), $\tau_{0}$ is attempt-to-escape from a trap (in $s$), $\Delta z$ is the inter-trap distance, $E_{C}-E_{F}$ is equivalent to $E_{\text{Amor}}$, and other symbols have their usual meanings. For plot shown in Figure 2, using $E_{\text{Amor}}$ and approximating for $\Delta z =$ 10 nm and $\tau_{0}=\,10^{-14}\,s$, we extract $u_{\text{a}}$ as 37.40 nm, and $N_{t}$ = $588m^{-3}$. From these, $F_{\text{th}}$ is obtained. The field equation is then used to determine $u_{\text{a}}$ for any other resistance state, by using the variable $V_{\text{th}}$ unique to that state and the constant $F_{\text{th}}$. For the state-dependent drift measurements, the resistance vs time measurement was performed for $500\,s$ duration. For temperature-dependent drift measurements, each data point was extracted from five resistance vs time measurements, each of $900\,s$ duration, where after every such measurement the device was crystallized for $u_{\text{a}}$ extraction. By fitting resistance vs time data for each state with the resistance drift equation $\nu=\frac{log(\frac{R_{t}}{R_{0}})}{log(\frac{t}{t_{0}})}$, we extracted the drift coefficients. Here, $R_{\text{0}}$ is the device resistance at time $t_{0}$, and $t$ is the time elapsed after $t_{0}$. 

\section*{References}
\def\url#1{}


\begin{figure}[h!]
\centering
    \includegraphics[width=0.75\textwidth]{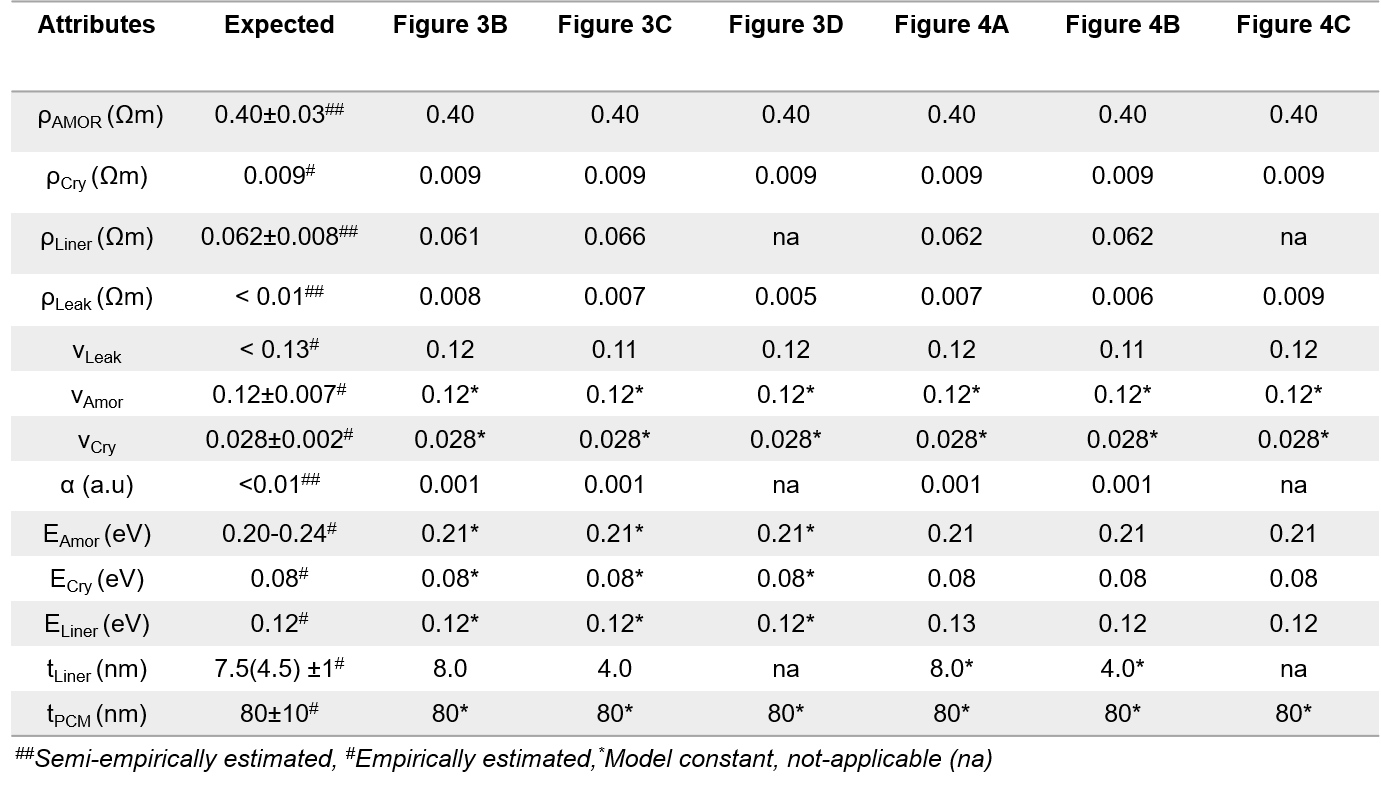}
    \caption*{\textbf{Table 1.} A listing of different material and geometrical parameters required by the device model. These parameters are obtained either empirically, semi-empirically or through fitting. The resistivites ($\rho$), drift coefficients ($\nu$) and film thickness ($t$) are values at 300 K.}
    \label{fig:6}
\end{figure} 

\end{document}